\documentclass[10pt,conference]{IEEEtran}
\IEEEoverridecommandlockouts
\usepackage[caption=false]{subfig}
\usepackage[pdftex]{graphicx} 
\usepackage[T1]{fontenc}
\usepackage{cite}
\usepackage{amsmath,amssymb,amsfonts}
\usepackage{algpseudocode, algorithm,algorithmicx}
\usepackage{braket}
\usepackage[pdftex]{graphicx} 
\usepackage[T1]{fontenc}
\usepackage{cite}
\usepackage{booktabs}
\usepackage{amsmath,amssymb,amsfonts}
\usepackage{algpseudocode, algorithm,algorithmicx}
\usepackage{braket}
\usepackage[symbol]{footmisc}

\DeclareGraphicsExtensions{.pdf,.jpeg,.png}

\usepackage{color}
\usepackage{soul}
\usepackage{cleveref}
\usepackage{nicefrac}

\crefname{figure}{Fig.}{Figs.}

\begin{document}
\title{Distributed quantum error correction based on hyperbolic Floquet codes\\
\thanks{This work was supported by Innovate UK grant 10074653.}}

\author{\IEEEauthorblockN{Evan Sutcliffe\IEEEauthorrefmark{1}\textsuperscript{$\dagger$}\textsuperscript{$\ddagger$}, Bhargavi Jonnadula\IEEEauthorrefmark{1}\textsuperscript{$\ddagger$},  Claire Le Gall\IEEEauthorrefmark{1}, Alexandra E. Moylett\IEEEauthorrefmark{1}\textsuperscript{$\ddagger$} and Coral M. Westoby\IEEEauthorrefmark{1}\textsuperscript{$\ddagger$}}
\IEEEauthorblockA{\IEEEauthorrefmark{1}\textit{Nu Quantum Ltd.}, Cambridge, United Kingdom \\
coral.westoby@nu-quantum.com}
}

\maketitle

\begin{abstract}
Quantum computing offers significant speedups, but the large number of physical qubits required for quantum error correction introduces engineering challenges for a monolithic architecture.
One solution is to distribute the logical quantum computation across multiple small quantum computers, with non-local operations enabled via distributed Bell states.
Previous investigations of distributed quantum error correction have largely focused on the surface code, which offers good error suppression but poor encoding rates, with each surface code instance only able to encode a single logical qubit.
In this work, we argue that hyperbolic Floquet codes are particularly well-suited to distributed quantum error correction for two reasons.
Firstly, their hyperbolic structure enables a high number of logical qubits to be stored efficiently.
Secondly, the fact that all measurements are between pairs of qubits means that each measurement only requires a single Bell state. Under the circuit-level noise model, we demonstrate through simulations that distributed hyperbolic Floquet codes offer good performance with achievable local and non-local fidelities of approximately $99.97\%$ and $99\%$, respectively. This shows that distributed quantum error correction is not only possible but also efficiently realisable.
\end{abstract}


\begingroup\renewcommand\thefootnote{$\dagger$}
\footnotetext{Current affiliation: \textit{Optical Networks Group}, \textit{Electronic \& Electrical Engineering Department}, \textit{University College London}, London, United Kingdom \\
evan.sutcliffe.20@ucl.ac.uk}
\endgroup

\begingroup\renewcommand\thefootnote{$\ddagger$}\footnotetext{These authors contributed equally.}
\endgroup

\section{Introduction}\label{introduction}

Quantum algorithms have the potential to solve problems such as factoring and simulating chemical systems exponentially faster than the best classical algorithms \cite{Montanaro2016}. Such
algorithms will require many more operations than can be performed
reliably on current quantum devices. As such, fault-tolerant quantum
computers will require quantum error correction (QEC), where many physical qubits are used to represent a smaller number of logical qubits, to store and process quantum information \cite{Roffe2019QECIntro}. These codes enable reliable operations in
the presence of noise, at the cost of requiring thousands or even millions of physical qubits for industrially relevant applications \cite{Reiher2017, Lee2021TensorHypercontraction, Gidney2021Factoring, Campbell2022HubbardModel, Blunt2022Perspective}. As such, it is desirable to find codes that can scale efficiently in their parameters $[[n, k, d]]$: the number of physical qubits $n$, the number of logical qubits $k$, and the number of correctable errors $d$ (``code distance''), should all grow at roughly the same rate.

The surface code is one of the best-known quantum error-correcting codes \cite{Kitaev2003ToricCode, Bravyi1998SurfaceCode} with parameters $[[n, 1, O(\sqrt{n})]]$. It is fault tolerant up to error rates of approximately $0.5\%$ \cite{Gidney2021honeycomb}, and can be implemented
with only nearest-neighbour interactions on a square array of qubits. These properties have enabled early demonstrations of quantum error correction on current monolithic quantum architectures, where a logical qubit has been shown to preserve quantum information up to  2.4 times longer than a physical qubit \cite{Google2023suppressing, Google2024BelowThreshold}. Efficiently implementing logical quantum operations on the surface code is also well-understood using techniques such as lattice surgery \cite{Horsman2012LatticeSurgery, Fowler2019LatticeSurgery, Litinski2019, Chamberland2022LatticeSurgery}. However, each instance of the surface code is only capable of storing a single logical qubit regardless of the number of physical qubits used. Thus while the surface code is able to correct large numbers of errors, it is not able to encode many logical qubits. More efficient encodings are possible by constructing a surface code on a closed (semi-)hyperbolic surface rather than a planar surface \cite{Breuckmann2017}, but checking for errors on these codes requires deeper circuits and non-planar connectivity.

To reach the millions of physical qubits required by industrially relevant applications, a promising approach is a \emph{distributed} architecture, where fixed size quantum processing units are interconnected by quantum channels. By using these channels to share entanglement between processors, the quantum circuit required to implement the QEC code can be distributed across the network \cite{Caleffi2022}. Furthermore, if these quantum channels are used to generate Bell states, then qubit teleportation or two-qubit gates can be performed using a shared Bell state, local operations, and classical communication \cite{Bennett1996}.

Some approaches to distributed quantum computation consider a setup in which each quantum processor holds a separate surface code patch \cite{Ramette2024, Singh2024modulararchitecturesentanglementschemes, Sinclair2024OpticalInterconnectsNeutralAtoms}. In this method, the network is only used for multi-qubit logical operations. An unwanted side-effect of this is that logical errors can occur along the edge of each quantum processing unit. This is problematic as errors are more likely to occur on communication qubits---physical qubits that are used to create the noisier non-local entanglement.

An alternative approach is to consider a fully distributed design, where each device holds only a moderate number of qubits, which together operate a single code. There are three main advantages of such a design. Firstly, it enhances the quantum resources available due to the modular design of small quantum devices. This implies that the total number of qubits in the system can be scaled without increasing the complexity of the individual quantum processors. Secondly, it benefits from non-planar connectivity through quantum channels, which allows interaction with qubits that are not immediate neighbours to each other. This capability opens up avenues for implementing error correction codes that encode more logical qubits for the same number of physical qubits, which is the primary focus of this paper. Lastly, the effective code distance is not restricted by the size of each processor. Furthermore, by relaxing the requirements for the number of qubits per device, a distributed fault-tolerant quantum computer can be constructed using quantum processor designs that balance the fidelity of operations with qubit count.

So far, proposed fully distributed architectures have focused on the surface code. A distributed surface code requires that Greenberger–Horne–Zeilinger (GHZ) states be shared between processors to perform weight-4 stabilisers non-locally \cite{Nickerson2013,deBone2024thresholds}. Such GHZ states are generated from multiple Bell states and are assumed to require multiple levels of entanglement distillation to achieve sufficiently high fidelity, thus requiring a high generation rate for non-local Bell states.

Another way to more efficiently encode quantum information is through using quantum low-density parity check (qLDPC) codes, which are inspired by classical LDPC codes \cite{Breuckmann2021}. It has been shown that such codes exist with asymptotically good parameters, meaning that the number of logical qubits and the code distance both scale linearly with the number of physical qubits \cite{Panteleev2022AsymptoticallyGoodQLDPC}. Small examples of quantum LDPC codes have also been explicitly developed with high pseudo-thresholds and encoding rates \cite{Bravyi2024}. However, these codes have high connectivity requirements for all qubits, making them challenging to implement in a monolithic device with limited long-range connectivity \cite{Berthusen2025LocalQLDPC}. The high connectivity requirements also mean that implementing good quantum LDPC codes in a distributed setting will put high demands on network throughput.

Similarly to surface codes, Floquet codes are quantum error-correcting codes which are defined on a topological surface \cite{Hastings2021dynamically}. Rather than statically encoding logical information as the codes described above do, Floquet codes dynamically encode logical information, allowing the information to evolve in such a way that the original information can still be retrieved.
A result of this dynamic encoding is that Floquet codes can check for errors using a series of pair-wise qubit measurements. Because each measurement is only between a pair of qubits, only a single Bell state is needed to implement a non-local measurement. 

Like the surface code, specific Floquet code constructions have been derived from closed (semi-)hyperbolic surfaces \cite{Higgott2023constructions,Fahimniya2024}. The code parameters of a hyperbolic Floquet code are $[[n, O(n), O(\log(n))]]$, which implies that the number of encoded logical qubits scales linearly with the number of physical qubits and the distance scales logarithmically with the number of physical qubits. Furthermore, the distance of these codes can be increased from $O(\log(n))$ to $O(\sqrt{n})$ while preserving the number of logical qubits through a process called fine-graining \cite{Higgott2023constructions}, where planar lattices are inserted to produce a semi-hyperbolic Floquet code. After a fine-graining level $f$ (see \Cref{sec:hyperbolic_semi-hyperbolic}), the initial code parameters $[[n, k, d]]$ change approximately to $[[f^2n, k, fd]]$. For certain physical error rates, such codes have been shown to require fewer physical qubits compared to using multiple copies of the planar surface code; see \cite{Higgott2023constructions}. Similarly to their (semi-)hyperbolic surface code counterparts, (semi-)hyperbolic Floquet codes require non-local connections to implement a closed surface. However, because Floquet codes only rely on weight-two measurements, they use simpler syndrome extraction circuits than (semi-)hyperbolic surface codes, with each non-local measurement requiring only a single Bell state shared over a quantum network.

\begin{figure}
    \centering
    \includegraphics[width=\linewidth]{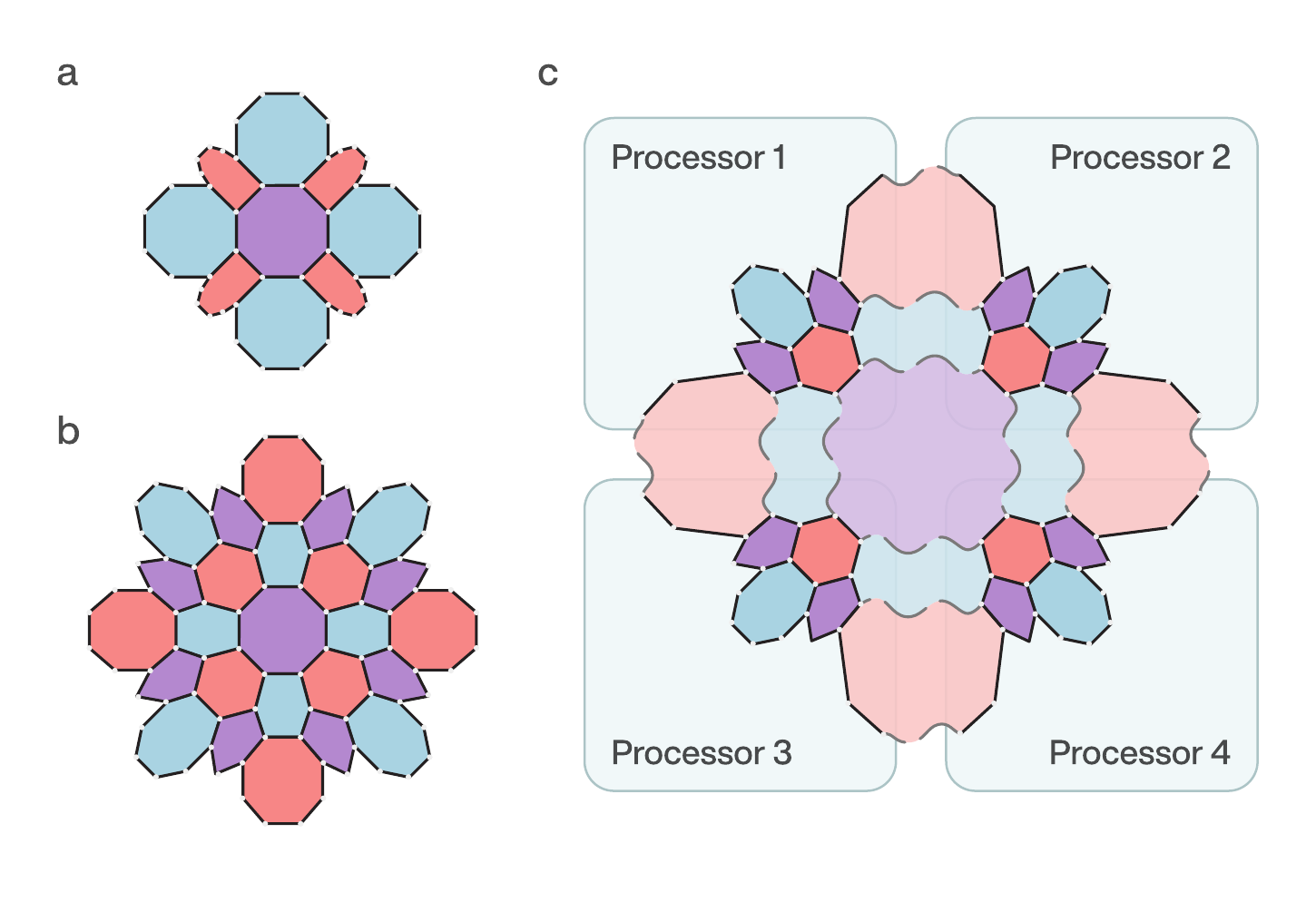}
    \caption{Summary of our approach for distributed quantum error correction. (a) A segment of a hyperbolic Floquet code defined on an octagonal lattice \cite{Higgott2023constructions, Fahimniya2024}. Each vertex represents a physical qubit, with edges representing two-qubit measurements and faces representing stabiliser measurements formed from the surrounding edge measurements. Further details on these codes are provided in \Cref{sec:Methods}. (b) Fine-graining approach is applied in order to produce a semi-hyperbolic code with a higher distance \cite{Higgott2023constructions}. (c) The higher-distance code is distributed across multiple quantum processors of a fixed size. Non-local measurements, denoted by grey lines, are implemented using distributed Bell states, which are generated using the architecture outlined in \Cref{sec:architecture}. \label{fig:summary}}
\end{figure}

As semi-hyperbolic Floquet codes provide a family of high-rate, locally planar, pair measurement codes of varying distance, we use them in this work to investigate the feasibility of distributed quantum error correction. 

Starting with a hyperbolic Floquet code (part of which is shown in \Cref{fig:summary}a), we use the fine-graining approach from \cite{Higgott2023constructions} to increase the distance of the code (\Cref{fig:summary}b). Finally, we distribute this code across multiple quantum processors of a fixed size, as shown in \Cref{fig:summary}c. Through simulations, we show that these codes are fault-tolerant under experimentally feasible numbers for both local and non-local circuit noise, and derive resource requirements for a quantum memory experiment of equivalent length to a million logical operations (``MegaQuOp'') \cite{Saha2024, Loschnauer2024}. This paves the way for implementing efficient quantum error-correcting codes in a distributed setting.

The rest of this manuscript is arranged as follows. In \Cref{sec:architecture}, we describe an illustrative architecture for distributed quantum error correction using non-local Bell states. We then provide further details on Floquet codes in \Cref{sec:Methods}, before discussing our simulation results in \Cref{sec:results}. Finally, we conclude in \Cref{sec:conclusions-and-further-work} with some open questions.

\section{Illustrating a networking architecture for distributed quantum error correction}
\label{sec:architecture}

\begin{figure}
    \centering
    \includegraphics[width=\linewidth]{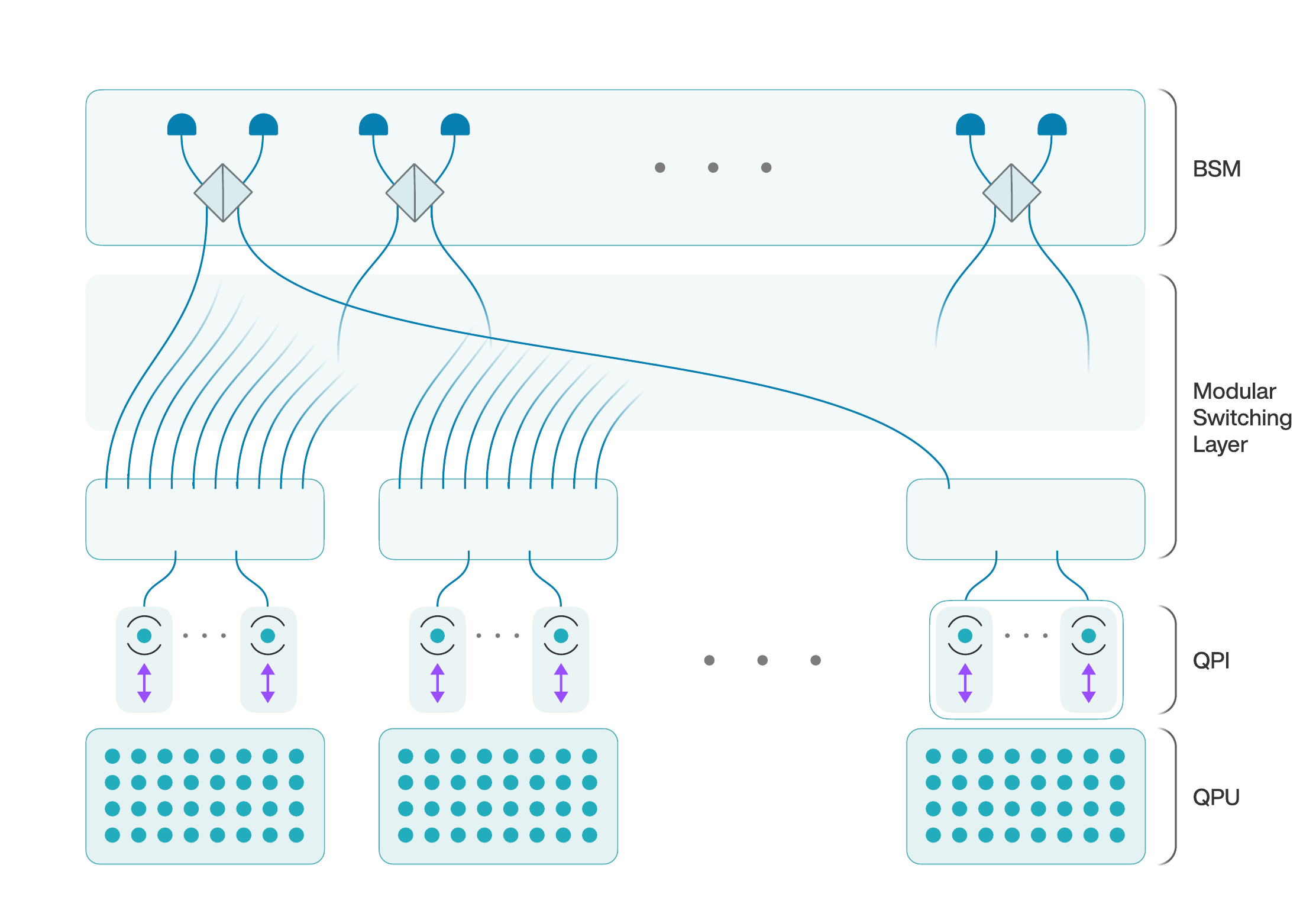}
    \caption{An overview of the networked quantum architecture, comprising quantum processing units (QPUs), qubit-photon interfaces (QPIs), photonic switches, and photonic Bell-State Measurement stations (BSM). Each QPU is interconnected with QPIs, where spin-photon entangled states are generated. The QPI-photons feed into a photonic switch, routing any of them to any of the switch outputs. The photon paths terminate at a BSM, where successful two-photon heralds project the QPI qubits into Bell states that can be used as remote entanglement links.
    \label{fig:architecture}}
\end{figure}

We will now give an example of an architecture that fulfils our need to generate non-local Bell states across multiple quantum processing units (QPUs) and efficiently distribute the QEC code.  This architecture is modular, meaning the code-size (or surface) is increased by using more identical QPUs and networking elements. A code which is locally-planar within the bounds of each QPU can be distributed across the entire system in such a way that the QPU connectivity (defined as the number of non-local links per QPU) stays constant. \Cref{fig:summary}c illustrates how this approach can be applied to a (semi-)hyperbolic Floquet code. Within a (semi-)hyperbolic Floquet code, a larger code size simultaneously increases the number of logical qubits \textit{and} improves the error suppression. Hence our illustrative architecture is able to support code scaling to arbitrary code sizes by using more QPUs, each of which feature an identical fixed number of qubits and non-local connections.

\Cref{fig:architecture} shows our architecture, which consists of four main components. First, the QPUs, which host computational qubits and perform local quantum operations\cite{AQT-QPU,Quantinuum-QPU}. Second, the qubit-photon interfaces (QPIs), which serve to entangle a QPI qubit and a photon\cite{InnsbruckQPI, SussexQPI}. Third, the switching layer, which provides a means to multiplex any number of QPIs and route photons towards photonic Bell-State-Measurement stations (BSMs), which are the fourth component of our architecture \cite{HarvardSwitch}. These BSMs perform heralded entanglement swapping across pairs of QPIs \cite{Stephenson2020HighRateEntanglement}. This architecture supports the sparse non-planar qubit connectivity graph required for high-rate quantum error correction codes \cite{baspin_quantifying_2022}.

Once entanglement has been established between the two QPI qubits, we need to interact these qubits with computational qubit(s) from their respective QPUs. This is possible in some solid-state platforms through hyperfine or dipolar interactions\cite{Bradley-QPI-QPU} and can be achieved in atomic platforms using shuttling across the QPI and QPU regions \cite{Akhtar2023,OrozcoRuiz2023}. The rate of remote entanglement generation will limit the rate at which non-local checks can be measured. The probability of generating a non-local Bell state via entanglement swapping is $\nicefrac{\eta^2}{2}$ where $\eta$ is photonic quantum efficiency from QPI to the detector, indicating that efficient and/or low loss components will support higher non-local entanglement rates and therefore typically improve quantum error correction performance. 

QPI qubits are only used as auxiliary qubits when implementing an error-correcting code. This allows us to mitigate the effects of photon loss by repeating non-local entanglement attempts until successful, ensuring photon loss does not contribute to infidelity to the first order. By removing this leading source of error, high fidelity non-local Bell states can be provided. Non-local Bell state fidelities of $97\%$  have been demonstrated with trapped-ion qubits and fidelities $>99\%$ are foreseen \cite{Saha2024}. Local gate fidelities of 99.99\%\cite{Loschnauer2024} have been demonstrated, showing that local fidelity should be expected to exceed remote Bell state fidelity. Non-local link fidelity exceeding the code threshold is needed to avoid the overhead of entanglement distillation, but nevertheless, higher noise on the non-local links must be tolerated.

\section{Quantum error correction with hyperbolic Floquet Codes}\label{sec:Methods}
In this section, we discuss a method for \emph{distributed quantum error correction}. We consider a large quantum error-correcting code that is distributed across many small quantum devices, enabling interactions between qubits that are not nearest neighbours. These connections are not scalable in a single monolithic device using on-device resources but are feasible using photonic interconnects for quantum networking as discussed in \Cref{sec:architecture}. 

For our distributed architecture, we employ hyperbolic Floquet quantum codes \cite{Higgott2023constructions, Fahimniya2024} to implement a quantum memory. These are a family of codes that use the tessellation of closed negatively curved planes to achieve both minimum-weight check operators and a constant encoding rate of logical to physical qubits. The large surface area of a hyperbolic surface allows a finite encoding rate due to the Gauss-Bonnet theorem \cite{Farb2011Primer}, which is a key feature that makes hyperbolic codes competitive with other known families of quantum error correction codes, such as the quantum LDPC codes \cite{Breuckmann2021, Panteleev2022AsymptoticallyGoodQLDPC, Bravyi2024}. We will now briefly discuss the structure of these codes and then move on to implementing them in a distributed architecture.

\subsection{Hyperbolic and semi-hyperbolic codes}
\label{sec:hyperbolic_semi-hyperbolic}
The hyperbolic code can be represented as a graph \hbox{$G = (V, E, F)$} where $V$, $E$, and $F$ are the sets of vertices, edges, and faces, respectively. Each face in $F$ is a polygon that tessellates the surface. The code requires that these faces are 3-colorable and the vertices are tri-valent, which calls for even-sided polygons. Data qubits live on the vertices $v\in V$, while the edges $e\in E$ connecting these vertices represent weight-two check operators. Each vertex has degree 3, so each qubit is connected to three other qubits. Each edge can have a Pauli check operator of the form $X\otimes X$, $Y\otimes Y$, and $Z\otimes Z$\footnote{It is worth noting that there are other instances of Floquet codes which are defined on the same lattice but implement measurements differently\cite{Hastings2021dynamically, Davydova2023floquet, Kesselring2024}.}. The colour of each edge can be determined from the colour of the faces it connects \cite{Hastings2021dynamically}. The lattice is closed using periodic boundary conditions, resulting in a genus $g$ surface. We indicate the lattice using Sch\"{a}fli symbols, for example, a lattice of the form $\{8, 3\}$ means that 3 octagons meet at each vertex. An $\{8, 3\}$ hyperbolic lattice with $n$ vertices has genus $g=n/16 + 1$. For detailed construction of hyperbolic lattices, see \cite{Foster1998, Higgott2023constructions, Fahimniya2024}. For the purpose of this article, the graphs are taken from \cite{Foster2020}.

\begin{figure*}
\centering
  {\includegraphics[width=\linewidth]{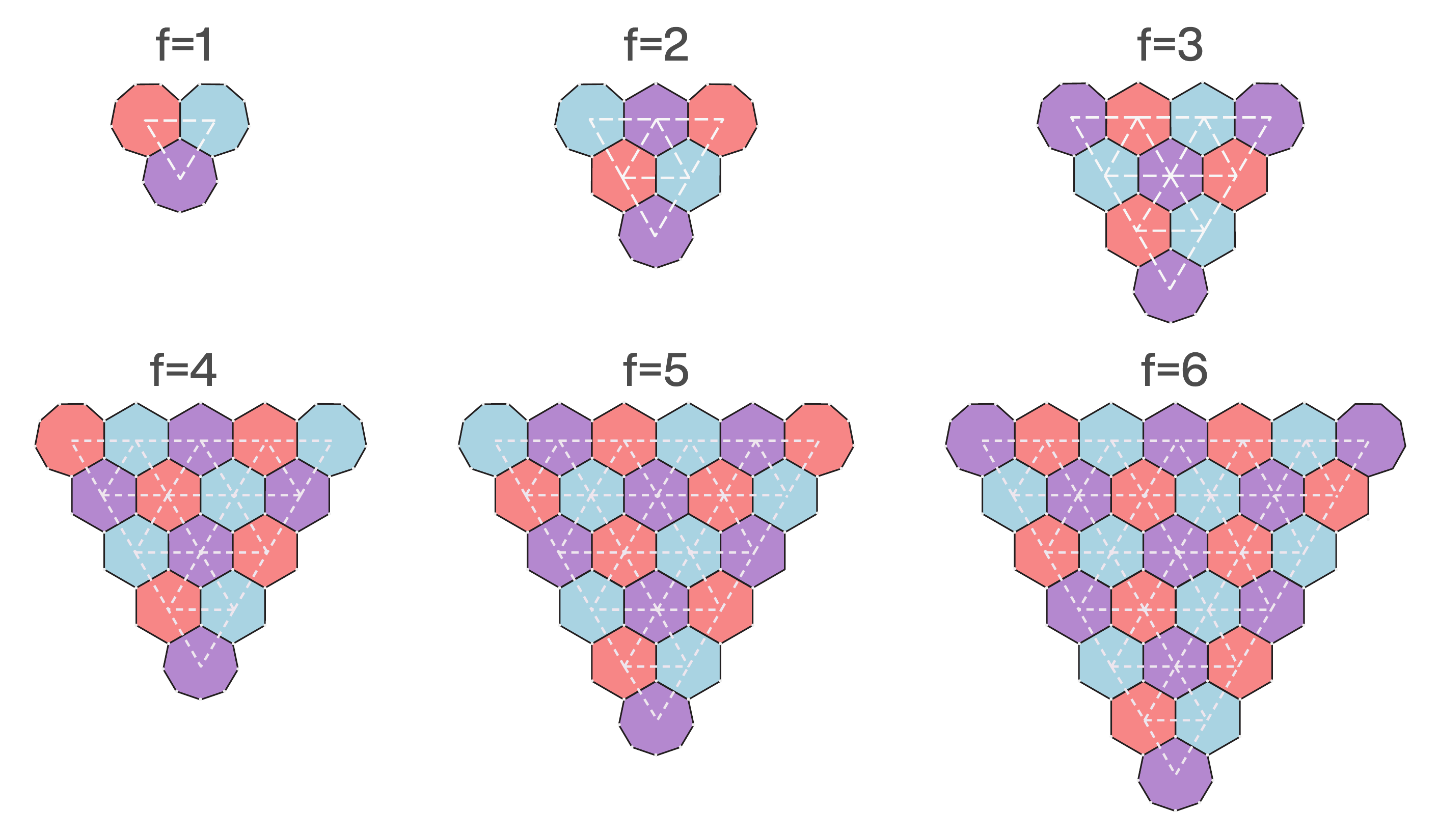}}
\caption{Applying the fine-graining procedure from \cite{Higgott2023constructions} to a vertex in an $\{8, 3\}$ tiling. Fine-graining level $f=1$ corresponds to the original lattice, with higher levels corresponding to increasing the size of the hexagonal sub-lattice to increase the code distance. This requires a re-colouring of the graph to accommodate the additional faces. White dashed lines indicate the dual of the graph and show how fine-graining corresponds to tiling the face of the dual graph with a triangular sub-graph.
\label{fig:finegraining}}
\end{figure*}

Consider a graph with $n$ vertices (or data qubits) and genus $g$. The number of logical qubits that can be encoded is denoted by $k=2g$ and the encoding rate of a code is denoted by $k/n=2g/n$. As mentioned earlier, the hyperbolic codes considered have efficient encoding rates but only achieve logarithmic scaling for code distance $d$. For example, the $\{8, 3\}$ hyperbolic lattice has genus $g=n/16+1$. Thus, the encoding rate is $k/n = 1/8+2/n$, which is constant in the limit $n\rightarrow\infty$, whereas the distance $d$ grows as $\log(n)$\cite{Fahimniya2024}. Using the fine-graining procedure in \cite{Higgott2023constructions}, the rate-distance trade-off can be adjusted in order to improve the distance at the expense of adding more qubits. Fine-graining a hyperbolic code adds planar patches to its surface, and hence the generated codes families are termed \emph{semi-hyperbolic} \cite{Higgott2023constructions,Breuckmann2017}. Since fine-graining preserves the topological properties of the lattice, the number of encoded logical qubits remains unchanged, while the distance can be increased by at most $O(\sqrt{n})$. \Cref{fig:finegraining} shows the fine-graining process in a small area of an $\{8,3\}$ octagonal tiling. To perform fine-graining, we begin with the dual of the lattice. In the dual lattice, the vertices are located on the faces of the initial lattice, and the edges are drawn between them. As a result, qubits that are on vertices in the original lattice are represented by triangles in the dual. Fine-graining involves further triangulating each of these triangles, i.e., each triangle in the dual graph is replaced with a triangle-bounded interior-triangular lattice of size $f$. The level of fine-graining can be increased by increasing $f$. The semi-hyperbolic lattice is obtained by taking the dual of this fine-grained lattice. In the case of $\{8,3\}$ tiling, the semi-hyperbolic lattice consists of both hexagons and octagons. Regarding the nomenclature, we represent hyperbolic codes as ``Hx'' where x is the number of data qubits. Semi-hyperbolic codes are named ``Hx-fy'', where Hx indicates the base hyperbolic code with x data qubits, and fy signifies fine-graining level y.

\subsection{Quantum code partitioning for modular
embedding}\label{ssec:graph-partitioning}

To implement the error correction code on a distributed quantum computer, we partition the graph corresponding to the semi-hyperbolic code across multiple small quantum units. Each quantum unit has a finite number of physical qubits, such as trapped ions or neutral atoms. The physical qubits constitute both data qubits from the error correction code and auxiliary qubits for syndrome extraction. The units are connected through the networking architecture discussed in \Cref{sec:architecture}.

The graph partitioning is done as follows. For a code represented by a graph $G$, it is divided into $N$ clusters $\{K_0,K_1,...,K_N\}$. Each cluster $K_i=(E_i,V_i)$ serves as a single quantum processing unit, composed of a set of vertices $V_i$ and edges $E_i$. The vertices $V_i$ consist of data qubits connected by the set of edges $E_i$. The edges $E_i$ correspond to weight-two checks, which we call local weight-two checks as they lie within the cluster $K_i$. For a partition satisfying $\bigcup_{i=0}^N V_i = V$, the non-local edges $E^\prime = E - \bigcup_{i=0}^N E_i$ connecting different quantum units correspond to non-local weight-two checks. Partitioning $G$ is performed using a heuristic approach called recursive spectral bisection\cite{Newman2018}. In our simulations, we ran recursive partitioning until each cluster met certain conditions. First, we need $\frac{3}{2}|V_i| < n_{\text{QPU}}$ where $n_{\text{QPU}}$ is the maximum number of physical computational qubits that can be accommodated within a quantum unit. We consider a fixed node size $n_{\text{QPU}}=32$, where 21 qubits are used for code data, and the remaining are reserved for check measurements. Additional qubit sites are needed to prepare the heralded
non-local Bell states. These sites are within the QPIs, and QPI qubits discussed in \Cref{sec:architecture} are not counted as computational qubits.  Second, as the QPUs are much smaller than the code, the subgraph on each quantum unit is planar. This leads to the majority of operations within a QPU being short-range, where computational qubits require planar connectivity to other computational qubits, enriched with longer-range connections to a communication qubit responsible for a portion of the subgraph boundary.

The weight-two checks within a cluster can be performed using local operations. The non-local weight-two checks can be performed using a single heralded shared Bell state and local operations. Thus, having a partitioned graph with a minimum number of non-local edges is beneficial as it requires fewer Bell states to perform error correction. We execute the non-local checks with a higher noise rate and accumulate a system-wide QPU additional delay that expresses the expected idling of the system while waiting for Bell states to be heralded.

In a physical implementation, we would use a cut-off wait time for the Bell state. In a Floquet code, a plaquette-forming round comprises three sub-rounds, namely $X\otimes X$, $Y\otimes Y$, and $Z\otimes Z$ checks. Each sub-round can be constrained to the duration of the check circuit. If the wait time is shorter than the sub-round, we will carry out further checks. If the wait time exceeds the sub-round, we will mark it as an error and continue with the subsequent round. 

\Cref{fig:summary} shows a potential approach for distributing part of a semi-hyperbolic Floquet code across multiple QPUs. For this code, the maximum number of physical qubits in each cluster is 32, which is feasible using current state-of-the-art technology \cite{Decross2024}. The connectivity within each unit is planar, but a routed heralded Bell state can facilitate non-planar connectivity. Each edge crossing the boundary
of the component is fixed. Therefore the connectivity requirements do not vary over the lifetime of the error correction code for a quantum memory experiment. However, the connectivity might have to be modified for computation \cite{Breuckmann2017,Higgott2023constructions,Breuckmann2024fold}.

\section{Simulation results}\label{sec:results}

We numerically estimated the logical error rates for the family of hyperbolic and semi-hyperbolic Floquet codes described in \Cref{sec:Methods}. The circuits were constructed and simulated using Stim, a tool for simulating quantum error correction circuits \cite{Gidney2021stim}. Errors detected by the Floquet codes considered here can be decomposed into weight-two ``graph-like'' errors, meaning that decoding can be performed using Minimum Weight Perfect Matching (MWPM) \cite{Higgott2025Sparse}. Control system latency will be introduced as the code is distributed over many QPUs, but we assume that for datacenter-scale installations this can be accommodated within the operation timescales of atomic qubits. We used a circuit noise model which also accounts for the challenges of running a distributed quantum code across a network, details of which are provided in \Cref{ssec:noise-model}.

From these simulations we estimated pseudo-thresholds for a circuit noise model. Furthermore, we investigated the effect of non-uniform noise introduced by partitioning the code across multiple devices. Finally, we estimate the number of physical qubits required to provide a MegaQuOp quantum memory for our distributed hyperbolic Floquet codes.

\subsection{Noise model}\label{ssec:noise-model}

The development of quantum error correction aims to identify a specific process of operations to protect quantum information stored in a noisy environment. Quantum memory experiments are performed based on an error model to reflect potential real errors that may occur, ensuring that theoretical successes can be translated into practical applications.
In this work, we assume that the primary source of error stems from an uncontrolled environment or through the quantum network, rather than imprecise qubit gate tuning. Hence, the error models are defined in terms of depolarising channels, rather than Pauli error biases. We employ a modified circuit-level noise model as our primary error model.

In a distributed setting, non-local weight-two checks are performed using a shared heralded Bell state that serves as an auxiliary qubit(s) for error detection \cite{deBone2020}. It is reasonable to assume that the fidelities of these Bell states during initialization will not be comparable with the initialization fidelities of the single auxiliary qubits within the quantum processing unit. Additionally, we must consider the time delay associated with generating the Bell state. In the following, we outline a method to incorporate these noise parameters into the simulations.

In the standard circuit-level noise model controlled by a single error parameter $p_{\textrm{local}}$, each single-qubit operation (state preparation, gate, and reset) is followed by a single-qubit depolarising channel with error probability $p_{\textrm{local}}$. Each two-qubit gate is followed by a two-qubit depolarising channel with error probability $p_{\textrm{local}}$. We apply a depolarising channel with the $p_\textrm{local}$ error probability prior to a qubit measurement. There is a chance that a measurement result is incorrectly recorded, which occurs with probability $p_{\textrm{local}}$. In addition to standard assumptions, we also consider noisy Bell state generation for non-local checks. We assume that a depolarising noise with probability $p_{\textrm{non-local}}$ is applied after the creation of the Bell state.
 
In each of the $X\otimes X$, $Y\otimes Y$, or $Z\otimes Z$ plaquette-forming rounds, the weight-two checks are distributed across the network according to the partitioning discussed in \Cref{ssec:graph-partitioning}. We then wait for the formation of Bell states corresponding to non-local checks to complete the plaquette-forming round. Generating heralded non-local entanglement is inherently stochastic, necessitating multiple attempts to successfully generate a Bell state \cite{Saha2024}. We suppose that number of attempts required to initialise a Bell state is 5 gate cycles, as a consequence of a assumed 20\% success rate of the cavity enhanced and heralded remote bell pair generation scheme\cite{Palacios-Berraquero2024Distributed}. We assume that the data qubits remain idle while waiting to be involved in an operation during which they experience decoherence. During this time, they encounter $t$ layers of depolarizing noise with probability $p_{\textrm{local}}$, where $t$ represents the waiting time for each qubit which can be kept track. We perform simulations by repeating detector rounds, where a detector round comprises 2 plaquette-forming rounds (6 sub-rounds).

\Cref{table:noise_model} provides our full noise model in terms of the gates available in Stim \cite{Gidney2021stim}. We assume negligible cross-talk or correlated errors, i.e., applying a gate on qubit 1 does not change the errors experienced by qubit 2.

\begin{table}
  \caption{Error probabilities of gate types under the circuit noise model}
  \label{table:noise_model}
  \centering
  \begin{tabular}{|c|c|}
    \hline
    \textbf{Gate type} & \textbf{Error parameter} \\ \hline
    Single-qubit (prepare, gate, reset, measure) & $p_{\textrm{local}}$ \\ \hline
    Two-qubit gate & $p_{\textrm{local}}$ \\ \hline
    Remote Bell state prepare & $p_{\textrm{non-local}}$ \\ \hline
    Decoherence (per circuit step) & $p_{\textrm{local}}$ \\ \hline
  \end{tabular}
\end{table}

\subsection{Pseudo-threshold}\label{error-threshold}

\begin{figure}
\centering
  {\includegraphics[width=\linewidth]{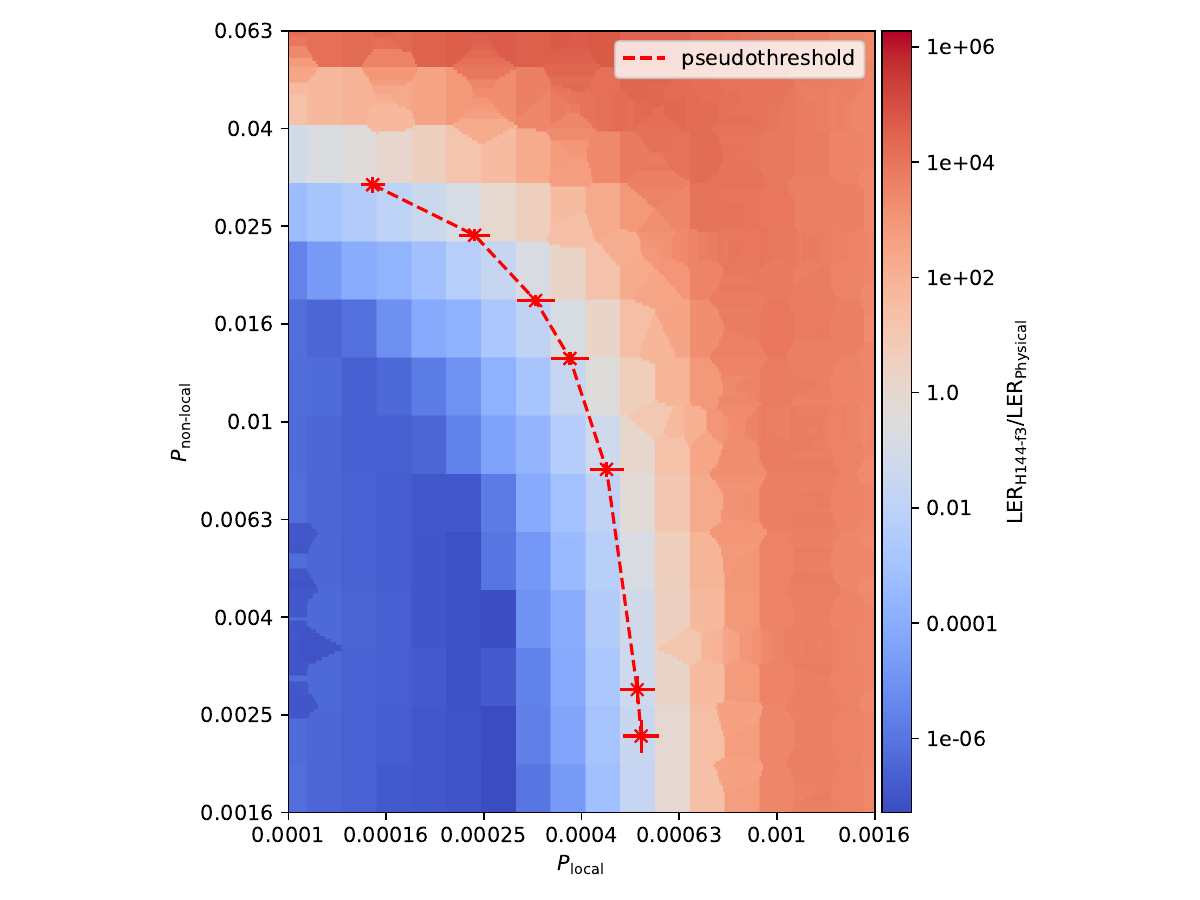}}
\caption{Pseudo-threshold for H144-f3. Logical error rates are shown for a single worst-case observable in the code H144-f3 (1296 data qubits) partitioned over 32 qubit devices for 12 detector rounds (72 colour rounds). Error bars in the threshold fitting are shown as solid red lines around threshold points. The ratio of logical to physical error rate is shown as a 2D color map. }
\label{fig:e5}
\end{figure}

The (pseudo)-threshold of a distributed quantum code should be understood as deriving from two factors - local and non-local noise processes. As such, we estimate the threshold curve in the 2D space of $p_{\textrm{local}}$ and $p_{\textrm{non-local}}$ errors.

Throughout the simulation results, we will look at the single logical Z observable with the highest logical error rate. The relative performance of each logical is weakly affected by the partitioning used. We use Stim \cite{Gidney2021stim} and PyMatching \cite{Higgott2025Sparse} to predict logical errors based on measurement results. 

We estimate the pseudo-threshold by comparing the relative error rates of H64-f3 and a single physical qubit. The physical qubit experiences a single-qubit error with probability $p_{\text{local}}$ for each detector round of the Floquet code. By comparing equivalent numbers of detector rounds, rather than time, we hope to find the computational pseudo-threshold where the logical observable has the same logical error rate (LER) as an observable on a single physical qubit experiencing $p_{\text{local}}$ depolarising error per detector round. This threshold is stricter than that for preserving logical information for a fixed wall clock time, as detector rounds are much longer than a single local gate time, but is useful for comparing the number of operations possible on a qubit in a fixed error budget.

The estimation process starts by performing a 512-point grid scan of the 2-D region defined by $p_{\textrm{local}} \in 10^{[-4, -2.5]}$ and $p_{\textrm{non-local}} \in 10^{[-3, -0.5]}$. We sample until either $512$ errors or $2\times 10^5$ total shots have been taken for each point. We fit a proxy Gaussian process to this data using \cite{gpyopt2016}. We use individual proxy models for the logical error rates of both the H144-f3 code and a physical qubit. We derive a difference model, defined as $|\log(p_{\textrm{encoded}})-\log(p_{\textrm{unencoded}})|$, where $p_{\textrm{encoded}}$ and $p_{\textrm{unencoded}}$ denote the probabilities of a logical error for a logical qubit encoded in the H144-f3 Floquet code and a single physical qubit, respectively. This function of each logical error probability will be zero at the pseudo-threshold and allows us to use robust search processes to find physical error rates on the pseudo-threshold.

We perform 7 steps of refinement sampling using local penalization with a batch size of 10 to refine the estimated pseudo-threshold. We perform this sampling within fixed "contexts", corresponding to points drawn from 2 linear scans over the $p_{\textrm{local}}$ and $p_{\textrm{non-local}}$ ranges. This ensures we find many minima, corresponding to points that cover the entire pseudo-threshold curve. The 2-D logical error rate landscape was interpolated onto a $800\times800$ point regular grid with SciPy's \texttt{griddata} function \cite{2020SciPy-NMeth} and smoothed using a Gaussian kernel with size 7.  Initial threshold estimates were drawn from this smoothed landscape. The interpolated landscape is plotted as the 2D colour map in \Cref{fig:e5}.

Starting from each initial estimate, we perform narrow sweeps to identify the pseudo-threshold with higher precision. These sweeps are performed until we have either $100$ errors or $4\times10^7$ shots. We sweep over the recorded pseudo-threshold in both the $p_{\textrm{local}}$ and $p_{\textrm{non-local}}$ axes for each point. We then perform a linear fit to both the physical and H144-f3 data from those sweeps, noting that the data is approximately linear due to the narrow sweep range. We take the point of intersection of those fits as the threshold. The error in the fits is shown as error bars on the threshold points.

The pseudo-threshold is shown in \Cref{fig:e5}. Red points indicate fitted locations of the pseudo-threshold, and the 2D color map corresponds to the error ratio between H144-f3 and a physical qubit for a quantum memory experiment. In the distributed setting, a Floquet encoded qubit can tolerate a higher $p_{\textrm{non-local}}$ error compared to $p_{\textrm{local}}$. At the highest tested local error rate, we find a pseudothreshold at $96.9\pm3.7\%$ remote bell state preparation fidelity and $99.99\pm0.02\%$ local fidelity. The pseudothreshold curve traces a smooth landscape through to $99.71\pm0.56\%$ remote bell state fidelity and a $99.95\pm0.14\%$ local fidelity. The intuitive reason for the higher noise tolerance of the non-local connections is that they represent a small fraction of the entanglement resources used in performing the parity checks.
Hence, when an error is detected, the decoder is more likely to associate it with the non-local link, which enhances the tolerance for non-local errors.

\subsection{Sub-threshold performance}\label{section:subth}

The practicality of a large-scale error corrected quantum computation is sensitive to the rate at which errors are suppressed for large codes in the early sub-threshold regime\cite{Google2021ExpSuppression}. This is characterised by the error suppression exponent $\Lambda$, which can be written as 
\begin{equation}
\label{eq:lambda}
    \epsilon_L = C \Lambda^{-(d+1)/2},
\end{equation}
where $\epsilon_L$ is the logical error rate per detector round where a detector round comprises 2 plaquette-forming rounds (6 sub-rounds), $d$ is the code distance, and $C$ is a fitting constant. $\epsilon_L$ is determined by expressing the overall logical error rate over a quantum memory experiment as

\begin{equation}\frac{1}{2} \left[1-( 1-2\epsilon_L )^{n_{\text{rounds}}}\right],\end{equation}

\noindent where $n_{\text{rounds}}$ is the number of detector rounds in the experiment.

We perform $\epsilon_L$ fits to codes derived from the H64, H144, and H400 base tilings. We sample H64-f$i$ for $i \in \{1,2,3,4\}$, and H144-f$j$ and H400-f$j$ for $j \in \{1,2,3\}$. 
We used the Stim function \verb|shortest_graphlike_error| to calculate the code distance. As the correct way to compute the distance of the hyperbolic codes is complicated by the dynamic support of the logical operators during the detector cycle, the distance for hyperbolic codes is not analytically known. The distance calculated by Stim is identical to $d(\text{H}n\text{-f}l) = \text{round}(\log(n)) \times l$ for all codes tested.
The logical error rates are the rates for the Z observable of the single worst performing logical supported in each code. The $\epsilon_L$ values derived from those fits are plotted in the right panel, and $\Lambda$ is found through a further fit as indicated in  \Cref{eq:lambda}. By sampling several base codes and further fine-graining levels of each base code, we observe by the quality of fit that the values of $\Lambda$ found are useful for predicting the performance of these codes under scaling through fine-graining.

\begin{figure}
    \centering
    \includegraphics[width=1\linewidth]{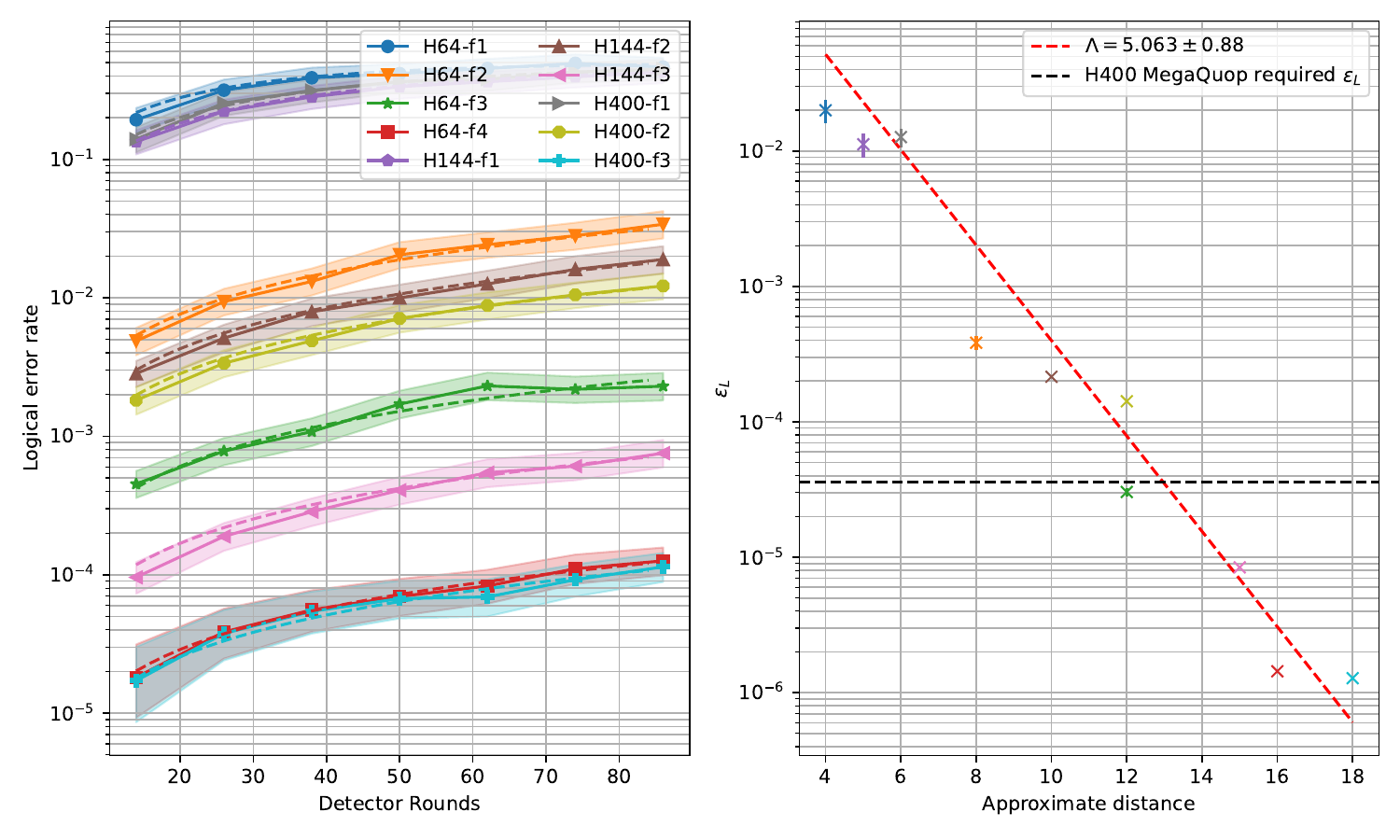}
    \caption{ Figures show (left) the logical error rate for code families with error rates fixed at $p_{\textrm{local}}=99.97\%$ and $p_{\textrm{non-local}}=99\%$. The sub-threshold error correction scaling $\Lambda$ (right) is calculated using \Cref{eq:lambda}. The data points in the right plot are derived from the error rate fits in the left plot.}
    \label{fig:lambda}
\end{figure}

For the error model under evaluation, $\Lambda = 5.06\pm0.88$ indicates the potential for a strong exponential suppression of logical error rates through modular scaling. H400 provides 52 logical qubits, and so in order to perform an identity MegaQuOp, that is, $10^6$ logical operations each requiring a detector round with a cumulative any-observable 50\% chance of failure, a system would require an observable-level success rate greater than $\exp(\log(0.5)/104) = 99.3\%$ after $10^6/52 = 19230$ detector rounds. 
This implies a failure rate of $3.5\times10^{-5}$ per detector round. This failure rate is shown on \Cref{fig:lambda} as a black horizontal line. At the tested error rates, H400-f3 would satisfy this requirement using 5400 qubits including measurement ancilla, in under 170 32-qubit devices equipped with 8 qubit-photon interfaces per device.

\section{Conclusion}\label{sec:conclusions-and-further-work}

We have shown that hyperbolic Floquet codes are a viable option for running quantum error-correcting codes over a network of many small quantum processors with near-term qubit numbers and error rates. These codes have efficient encoding rates and low weight stabilisers, making them tolerant to higher error rates between devices. Results from our numerical simulations show that the distributed hyperbolic code H144-f3 can tolerate non-local noise of up to 3\% without entanglement distillation. This shows how the codes can tolerate non-local noise up to an order of magnitude higher than the threshold for local noise. For a circuit noise model with feasible local ($99.97\%$) and non-local ($99\%$) error rates, we find that hyperbolic Floquet codes achieve good error suppression and feasible resource requirements for running a million quantum operations. From this we conclude that distributed error correction offers a viable, practically feasible, and efficient path for scaling quantum computation. This enables a complementary path to quantum utility alongside efforts to improve the number of physical qubits or qubit fidelities per device.

One area of further work is understanding how to perform logical gates on hyperbolic Floquet codes. For useful fault-tolerant quantum computation, the logical gates should have low execution complexity. It is known that logical operations can be implemented on planar Floquet codes through lattice surgery \cite{Haah2022BoundaryHoneycomb}. However, other techniques with lower overhead may exist, motivated by the availability of long-range operations in a distributed architecture involving photonic interconnects. Long-range operations might support approaches to universality that avoid the $O(d^3)$ space-time volume scaling seen for universal gate sets in strictly local settings. In principle, universal quantum gates are possible through braiding Anyon excitations, although this is only discussed with respect to Fibonacci codes\cite{Lavasani2019DehnHyperbolic}  or with 3-D extensions\cite{Davydova2024DACodes}. It is not yet known if these operations can extend to high rate codes, or how errors might propagate between logical gates. We also hope that an error-correction-aware graph partitioning process may improve on the partitioning scheme used in this work, possibly leading to improved error thresholds.

\section*{Acknowledgement}

We thank Marina Matijaca and MetaDeck for support with the design of \Cref{fig:summary,fig:architecture,fig:finegraining}. We thank our colleagues at Nu Quantum for helpful discussions, and thank Carmen Palacios-Berraquero for providing an environment where this work was possible.

\bibliographystyle{IEEEtran}
\bibliography{IEEEabrv,bib}

\end{document}